\documentstyle[aps,prd,twocolumn,epsf,epsfig]{revtex}
\newcommand \bea{\begin{eqnarray}}
\newcommand \eea{\end{eqnarray}}
\newcommand \beq{\begin{eqnarray}}
\newcommand \eeq{\end{eqnarray}}

\begin{document}
\twocolumn[\hsize\textwidth\columnwidth\hsize
\csname@twocolumnfalse%
\endcsname
\draft
\title{Influence of Induced Interactions on the Superfluid Transition
in Dilute Fermi Gases}
\author{H.\ Heiselberg$^1$, C.\ J.\ Pethick$^{1}$, H.\ Smith$^2$,
and L.\ Viverit$^2$}

\address{
{\setlength{\baselineskip}{18pt}
$^1$NORDITA, Blegdamsvej 17, DK-2100 Copenhagen \O, Denmark \\
$^2$\O rsted Laboratory, H. C. {\O}rsted Institute,
Universitetsparken 5, DK-2100 Copenhagen {\O}, Denmark} }

\maketitle

\begin{abstract}
\noindent We calculate the effects of induced interactions on the transition
temperature to the BCS state in dilute Fermi gases.  For a pure Fermi
system with 2 species having equal densities, the transition
temperature is suppressed by a factor $(4e)^{1/3}\approx 2.2$, and
for $\nu$ fermion species, the transition temperature is increased
by a factor $(4e)^{\nu/3-1}\approx 2.2^{\nu-3}$.  For mixtures of
fermions and bosons the  exchange of boson
density fluctuations gives rise to an attractive
interaction, and we estimate the increase
of  the transition temperature due to this effect.\\
\end{abstract}
\vskip1pc]

The study of possible superfluidity in dilute Fermi gases has a long
history stretching back to the years immediately after the development
of the BCS theory \cite{hove,emery,Gorkov}.  In recent years the
subject has received renewed attention for a variety of reasons.  The
first is that superfluidity of nucleons plays an important role in
theories of neutron stars and of finite nuclei.  The magnitudes of
superfluid gaps are necessary input for calculations of transport
properties and neutrino emission rates, and for modeling the glitches
observed in the rotational periods of neutron stars.  The second is
theoretical interest in how properties of a Fermi system
change as the strength of an attractive interaction is varied
\cite{leggett,nozieres,randeira}.  If the system has a two-body bound
state, it will behave at low densities as a collection of diatomic
molecules, while for weaker attraction it will behave as a BCS
superfluid.  A third reason is the possibility of observing the
transition to the BCS paired state in fermion alkali atom vapors in
traps \cite{stoof}.  Work on this topic has been
spurred on by the recent success in cooling dilute alkali atom vapors
to temperatures below the degeneracy temperature \cite{JILA}.

During the 1990's there have been a number  of papers that calculate the
transition temperature, or equivalently the zero temperature gap, of dilute
Fermi systems \cite{randeira,stoof,bertsch,clark}.
The basic physics of these papers amounts to summing ladder diagrams
with the bare fermion-fermion interaction,
and expressing the result in terms of the scattering length for
two-body scattering {\it in vacuo}.  For two spin components
with equal densities, these papers predict a transition
temperature $T_c$  given  by
\beq
    kT_{c0} =\frac{\gamma}{\pi}\frac{8}{e^2} E_F e^{1/N(0)U_0} \approx
  0.61 E_F e^{\pi/2k_Fa},
\label{Tc0}
\eeq
for weak coupling ($N(0)|U_0|\ll 1$). Here
$E_F =p_F^2/2m_F$ is the Fermi energy, $p_F = \hbar k_F$ is the Fermi
momentum, and $m_F$ is the fermion mass.  The quantity $N(0)=m_F
k_F/(2\pi^2\hbar^2)$ is the density of states at the Fermi surface and the
matrix element of the effective interaction is $U_0=4 \pi \hbar^2a/m_F$, 
where $a$ is the
scattering length for two-body scattering, which is negative for an attractive
interaction.  The number $\gamma$ is $e^C$, where $C\approx 0.577$ is Euler's
constant.

The above calculation does not take into account the effects of the medium on
the two-body interaction.  Physically these processes, which we refer to as
induced interactions, correspond to one fermion polarizing the medium, and a
second fermion is then influenced by this polarization.  This gives rise to an
interaction between fermions analogous to the phonon-induced attraction
responsible for pairing of electrons in metallic superconductors.  The effects
of such interactions on superfluid transition temperatures have been studied in
dense systems.  In liquid $^3$He they are responsible for the ABM state being
energetically favored close to the superfluid transition temperature, whereas
in their absence the BW state would be the equilibrium one \cite{woelfle}.  For
neutron matter calculations predict that they suppress the superfluid gap
significantly \cite{clark2,wambach,schulze} . The corresponding effect in a
dilute spin-$1/2$ Fermi gas was considered by Gorkov and
Melik-Barkhudarov, who found that the transition temperature was suppressed by
a factor $(4e)^{1/3}\approx 2.2$ compared with the result of Eq.~(\ref{Tc0})
\cite{Gorkov}.  In this paper we elucidate the physical origin of the
suppression, and we derive expressions
for the transition temperature for fermions with a larger spin degeneracy, and
for mixtures of fermions and bosons.

The transition temperature is determined by there being a solution to the
linearized equation for the gap $\Delta_{\bf p}$,
\begin{equation}
\Delta_{\bf p}=
-\sum_{{\bf p'}}^{}
U({\bf p},{\bf p'})
\frac{1-2f^0(\xi_{\bf p'})}{2\xi_{\bf p'}}\Delta_{\bf p'},
\label{gap1}
\end{equation}
where $\xi_{\bf p}
=p^2/2m_F -\mu$ is the energy of a fermion relative to the chemical
potential $\mu$, and $f^0(\xi)=[\exp(\xi/kT)+1]^{-1}$
is the Fermi distribution function.
The total interaction is given by
\begin{equation}
U({\bf p},{\bf p'}) =
U_{\rm bare}({\bf p},{\bf p'})+U_{\rm ind}({\bf p},{\bf p'}),
\end{equation}
where $U_{\rm bare}({\bf p},{\bf p'})$ is the bare two-body interaction, and
$U_{\rm ind}({\bf p},{\bf p'})$ is the induced interaction.
Solving the linearized gap equation is equivalent to finding the temperature at which the $T$-matrix
obtained by summing ladder diagrams for the repeated interaction of two
particles with equal
 and opposite momenta and with total energy $2\mu$ diverges.
It is convenient to eliminate the bare interaction in favor of
$T_0({\bf p},{\bf p'};2\mu)$, the $T$-matrix for scattering in free space of two fermions,
each with energy $\mu$.  This is given by
\bea
 \nonumber
T_0({\bf p},{\bf p'}; 2\mu) &=&   U_{\rm bare}({\bf p},{\bf p'})\\
&+& \sum_{{\bf p''}}
U_{\rm bare}({\bf p},{\bf p''})
\frac{1}{2\xi_{\bf p''}}  T_0({\bf p}'',{\bf p'};2\mu).   \label{T0}
\eea
Solving Eq.\ (\ref{T0}) for $U_{\rm bare}$
and inserting the result into (\ref{gap1}) one finds
\begin{eqnarray}
\Delta_{\bf p}=
\sum_{{\bf p'}}^{}
T_0({\bf p},{\bf p'};2\mu)
\frac{f^0(\xi_{\bf p'})}{\xi_{\bf p'}}\Delta_{\bf p'} \nonumber \\
-\sum_{{\bf p'}}U_{\rm ind}({\bf p},{\bf p'})
\frac{1-2f^0(\xi_{\bf p'})}{2\xi_{\bf p'}}\Delta_{\bf p'}\nonumber \\
+\sum_{{\bf p', p''}} T_0({\bf p},{\bf p''};2\mu)  \frac{1}{2\xi_{\bf p''}}
U_{\rm ind}({\bf p''},{\bf p'}) \frac{1-2f^0(\xi_{\bf p'})}{2\xi_{\bf
p'}}\Delta_{\bf p'}.
\label{gap2}
\end{eqnarray}
Let us first consider the transition temperature in the absence of induced
interactions.
The factor $f^0(\xi_{\bf p'})/\xi_{\bf p'}$ falls off rapidly for
momenta greater than the Fermi momentum.  According to the standard effective
range expansion, the on-shell $T$-matrix  for
small
$p$ and $p'$ is its value for
$p=p'=0$ and zero energy, $T_0(0,0;0)=U_0$, plus terms of order $p^2$.
The latter terms produce higher-order
contributions than those from the induced interaction, and will be neglected
here.  Thus in Eq.\ (\ref{gap2}) the gap may be put equal to a constant for
momenta less than or of order the Fermi momentum, and the temperature at which
there is a non-trivial solution to the equation is given by
\beq
1+U_0\int_0^{\infty}
\frac{p^2dp}{2\pi^2\hbar^3}\frac{1}{\xi}\frac{1}{e^{\xi/kT_{c0}}+1}=0.
\eeq
On performing the integration one arrives at Eq.\ (\ref{Tc0}).

We now include the effect of induced interactions, which we shall
assume to be small compared with the interaction of two fermions {\em
in vacuo}.  In Eq.\ (\ref{gap2}) the last term represents
 a final-state interaction.  The range of the induced interaction
is of order the spacing between fermions or,
for the phonon-induced interaction in mixtures,
the coherence length. Both these lengths are large compared with
the magnitude of the fermion-fermion scattering length,
and therefore in the region where the induced interaction is important
the wave function of two fermions in the presence of the bare interaction, which is equal to
$1-a/r$, is essentially equal to unity, and
final-state effects are negligible.
 One then finds that the transition temperature is given
to leading order in $U_{\rm ind}$ by
\beq
   kT_c =\frac{\gamma}{\pi}\frac{8}{e^2} E_F
   e^{1/N(0)(U_0+\bar{U}_{\rm ind})},    
\label{Tcind}
\eeq
where
\beq
   \bar{U}_{\rm ind} =\int_{-1}^{1} \frac{d \cos\theta}{2}
   U_{\rm ind}( p_F{\hat{\bf n}},p_F{\hat{\bf n}'}),  \label{Ubar}
\eeq
is the average of the induced interaction over the Fermi surface.  The angle
between the two momenta is denoted by $\theta$, i.\ e. $
{\hat{\bf n}}\cdot{\hat{\bf n}'}=\cos\theta$.  Note that the
frequency dependence of the induced interaction does not enter, only its
value at the Fermi surface.  When the induced interaction becomes comparable
to the bare interaction, the effects of the frequency dependence of the
interaction will become important.

\begin{figure}
\begin{center}
\psfig{file=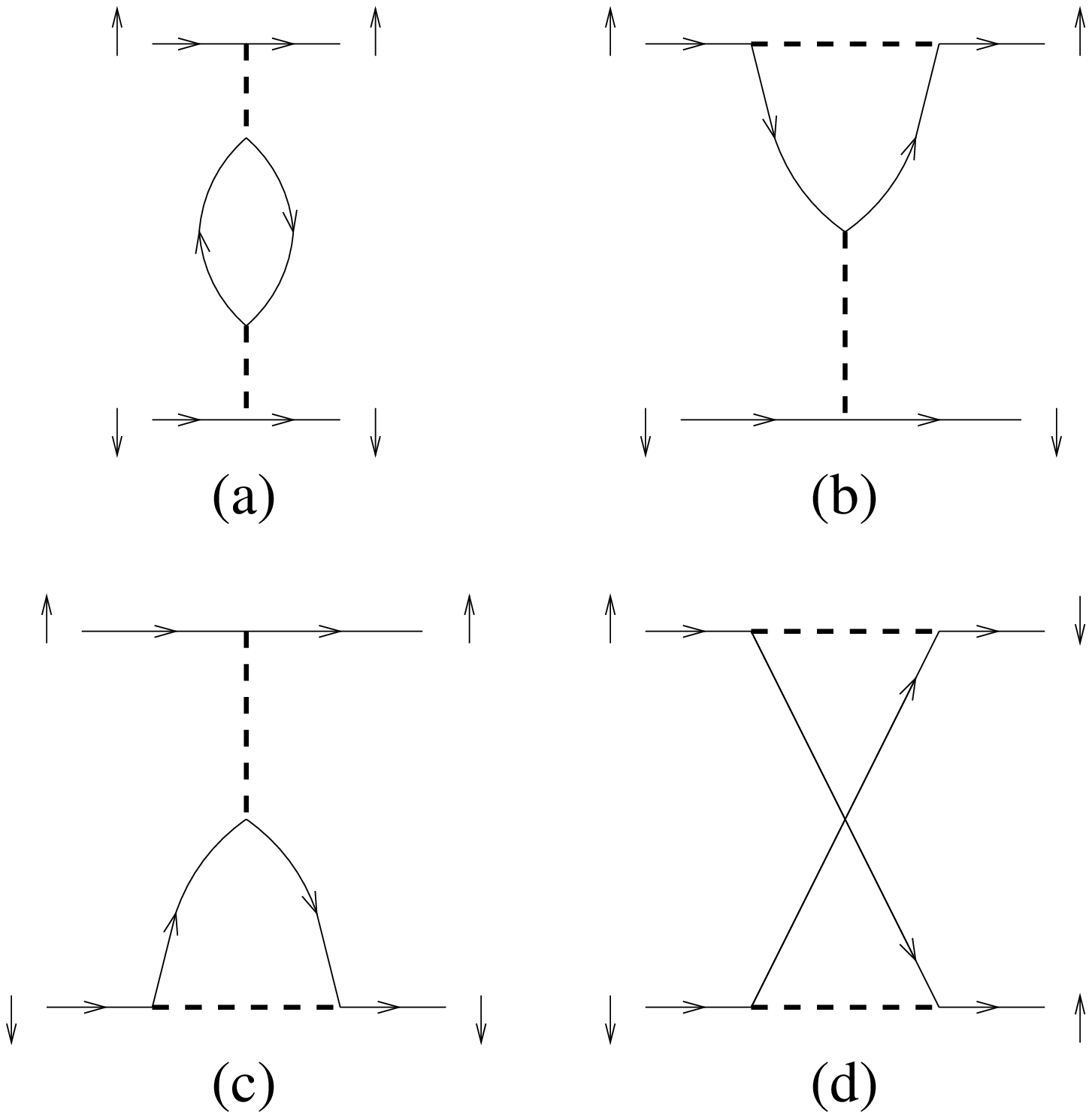,height=5.7cm,angle=-90}
\vspace{.2cm}
\begin{caption}
{Diagrams for the induced interactions between two fermions
in different internal states
to second order in the effective interaction.}
\end{caption}
\end{center}
\label{fig1}
\end{figure}

We now apply this result to Fermi systems, and we begin by considering a Fermi
gas with two internal degrees of freedom, denoted in Fig.\ 1 by up
and down arrows.
We shall take the densities of the two components to be
equal, since this gives the largest gap.  The leading contributions to the
induced interaction are represented diagrammatically 
in Fig.\ 1. One may ask why they should have any effect at all
in this limit, since they are formally of higher order in the density than the
leading term.  To understand this, consider the process shown in Fig.\ 1(a),
which is the screening of the interaction between two fermions by the other
fermions.  This is of order the density of states times the square of the
effective two-body interaction, and is therefore of
order $k_FaU_0$. If one alters the
effective
interaction in the expression (\ref{Tc0}) by an amount of relative order
$k_Fa$, it is easy to see that the gap is multiplied by a constant factor, as
the calculation of Gorkov and Melik-Barkhudarov demonstrated explicitly
\cite{Gorkov}.  For a contact interaction, the contributions of the diagrams
in Fig.\ 1 give

\beq
U_{\rm ind}( p_F{\hat{\bf n}},p_F{\hat{\bf n}'})=U_0^2 L(q)
\label{Uind2}
\eeq
where
\beq
L(q) = N(0)
\left(\frac{1}{2}
+\frac{(1-\eta^2)}{4\eta}\ln\left|\frac{1+\eta}{1-\eta} \right|
\right),
\label{lindhard}
\eeq
is the static Lindhard function and $\eta=q/2k_F$.
Here, $\hbar{\bf q}={\bf p'-p}$ is the momentum transfer in the
interaction which, as both particles are at the Fermi surface, is
related to the scattering angle $\theta$ by
$q^2=2k_F^2(1-\cos\theta)$.
Carrying out the integration over angles in Eq.\ (\ref{Ubar})
gives
\beq
   \bar{U}_{\rm ind} = U_0^2 N(0) \frac{1+2\ln2}{3} \,,
\label{Ubarf}
\eeq
and by insertion of this expression in Eq.\ (\ref{Tcind}) we find
\beq
   T_c=\frac{1}{(4e)^{1/3}}T_{c0},
\eeq
a result implicit in Ref.\ \cite{Gorkov}.
We therefore arrive at the striking conclusion that induced interactions reduce
the transition temperature by a factor $(4e)^{1/3} \approx 2.2$ even in the
low density limit.  It is interesting that calculations of the superfluid 
gap in
neutron matter at densities at which the low-density limit is inapplicable
indicate that induced interactions reduce the maximum of the gap as a
function of density by a comparable factor \cite{clark2,wambach,schulze},
but this is a coincidence.

The physics of the suppression is best examined by expressing the result in
terms of the amplitudes for exchange of density and spin fluctuations.  In
terms of these one finds for the interaction in the spin-singlet channel for
the pair of fermions
\beq
U_{\rm ind}(q)= [U_{s}(q) -3U_t(q)]/2,
\eeq
where $U_s$ and $U_t$ are the amplitudes for exchange of
spin-singlet particle-hole pairs (density fluctuations) and
spin-triplet ones (spin fluctuations) respectively.  The leading
contributions to the amplitudes are $U_s =-U_t= -U_0^2 L(q)$,
and therefore $U_{\rm ind}(q)=U_0^2L(q)$.  One way of expressing this
is to note that the contributions from the diagrams in Fig.\ 1(a)-(c),
which are due to exchange of spin and density fluctuations with
spin projection zero, cancel, leaving the contribution from exchange of
spin fluctuations with spin projection $\pm 1$, Fig.\ 1(d).  The
important process is therefore the suppression of superfluidity by
exchange of spin fluctuations, an effect familiar in metallic
superconductors and liquid $^3$He \cite{schrieffer}.

The result above may easily be generalized to $\nu$ fermion species.
For simplicity we assume that the two-body interaction is independent
of the species, and is diagonal in the species labels.  This is true
if the interaction is spin-independent and depends only on the
distance between the two fermions.  The contribution from the diagram
in Fig.\ 1(a) is proportional to $\nu$, because all species 
contribute to the closed loop, while the other diagrams in Fig.\ 1 are
independent of the number of species, since there are no fermion
loops.  Thus the contribution from all diagrams is the same as for two
species, except that the first term is multiplied by $\nu$, and their
sum is proportional to $\nu-3$.  The transition temperature is given by
\beq
T_c =   (4e)^{\nu/3-1} T_{c0},
\eeq
and therefore for four or more components the transition temperature is {\em
increased}, rather than decreased.

The results above are not directly applicable to dilute nuclear matter
because the nuclear force is strongly dependent on isospin: the
neutron-neutron scattering length is approximately $ -18.8$ fm, while
the neutron-proton one is $5.4$ fm.  As a consequence, the induced
interaction between two neutrons, say, due to excitation of neutron
particle-hole pairs is of order $(18.8/5.4)^2 \approx 12$ times
stronger than that due to excitation of proton pairs, and therefore
the latter contribution may be neglected to a first approximation.
The fact that induced interactions are important for pairing even at very 
low densities in
bulk systems indicates the need to investigate how significant these
effects are in the outer parts of finite nuclei.

\begin{figure}
\begin{center}
\psfig{file=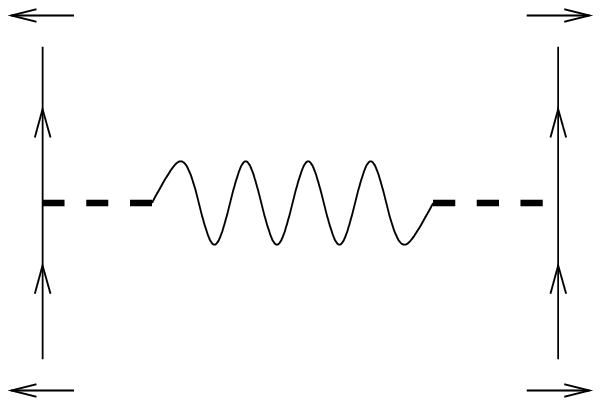,height=1.5cm,angle=-90}
\vspace{.2cm}
\begin{caption}
{Diagram for the phonon-induced interaction between two fermions. 
The dashed line is the fermion-boson interaction, and
the wavy line is a phonon in the boson gas.}
\end{caption}
\end{center}
\label{fig2}
\end{figure}

We now consider adding bosons of mass $m_B$ to a Fermi gas with two species.
The induced interaction between fermions then contains a
contribution due to the exchange of boson density fluctuations,
 analogous to the phonon-induced interaction between electrons in metals and
to
the induced interaction between $^3$He atoms in dilute solutions of $^3$He in
$^4$He. It is given by  \cite{viverit,bijlsma}
\beq
U_{\rm ind} = U_{\rm BF}^2 \chi(q, \omega),
\eeq
where $U_{BF}=4\pi\hbar^2a_{BF}/m_{BF}$
(with $m_{BF}=2m_F m_B/(m_F+m_B)$) is the effective interaction between a boson
and a fermion, and $\chi(q, \omega)$ is the density-density response function
for the bosons as a function of the energy transfer $\hbar\omega$.  For a
dilute Bose gas the response function at zero temperature is given by the
result of using the Bogoliubov approximation, and is

\begin{equation}
 \chi(q,\omega)=\frac{n_B\hbar^2 q^2/m_B}{(\hbar \omega)^2-
   \epsilon_q^0(\epsilon_q^0+2n_BU_{BB})},
\end{equation}
where $U_{BB}=4\pi\hbar^2a_{BB}/m_B$ is the boson-boson effective interaction
 and $\epsilon_q^0=\hbar^2q^2/2m_B$.
For particles at the Fermi surface the energy transfer is zero, and
therefore the induced interaction relevant for calculating the transition
temperature is
\begin{equation}
   U_{\rm ind}(q,0)=-\frac{U_{BF}^2}{U_{BB}}\frac{1}{1+(\hbar q/2m_B
s)^2},
\end{equation}
where $s=(n_B U_{BB}/m_B)^{1/2}$ is the sound velocity in the boson gas.
The average over the Fermi surface of the boson-induced interaction between
fermions that enters the equation for the gap is
\begin{equation}
\bar{U}_{\rm ind}=-\frac{U_{BF}^2}{U_{BB}}H(p_F/m_Bs),
\end{equation}
where $H(x)=\ln(1+x^2)/x^2$.
Observe that this result is independent of the density of fermions in
the limit of low fermion densities.  This is in contrast to the induced
interaction due to fermion-fermion interactions alone, which is
proportional to $p_F$.  If we again assume that $\bar{U}^{BF}_{\rm
ind}$ is small compared with the fermion-fermion interaction, the
transition temperature is given by Eq.\ (\ref{Tcind}), where the induced interaction
includes terms due to fermion-fermion interactions as well as fermion-boson 
interactions. In Fig.\ 3 this is
plotted as a function of the dimensionless quantity
$(U_{BF}^2/U_{BB}|U_{FF}|)H(p_F/m_Bs)$. (To make the notation uniform
we here denote the fermion-fermion interaction, which we
assume to be attractive, by $U_{FF}$ rather than $U_0$ as we did in
the earlier part of the Letter.) The parameter  
$U_{BF}^2/U_{BB}|U_{FF}|$ is a measure of the importance of the induced 
interactions compared with the direct one. Since the function $H$ decreases
monotonically from its maximum value of $1$ for
$x=0$, the largest relative effects are achieved for small
fermion densities.  As Fig.\ 3 shows, the increase in $T_c$ due 
to the presence of the bosons can be considerable.

\begin{figure}
\begin{center}
\psfig{file=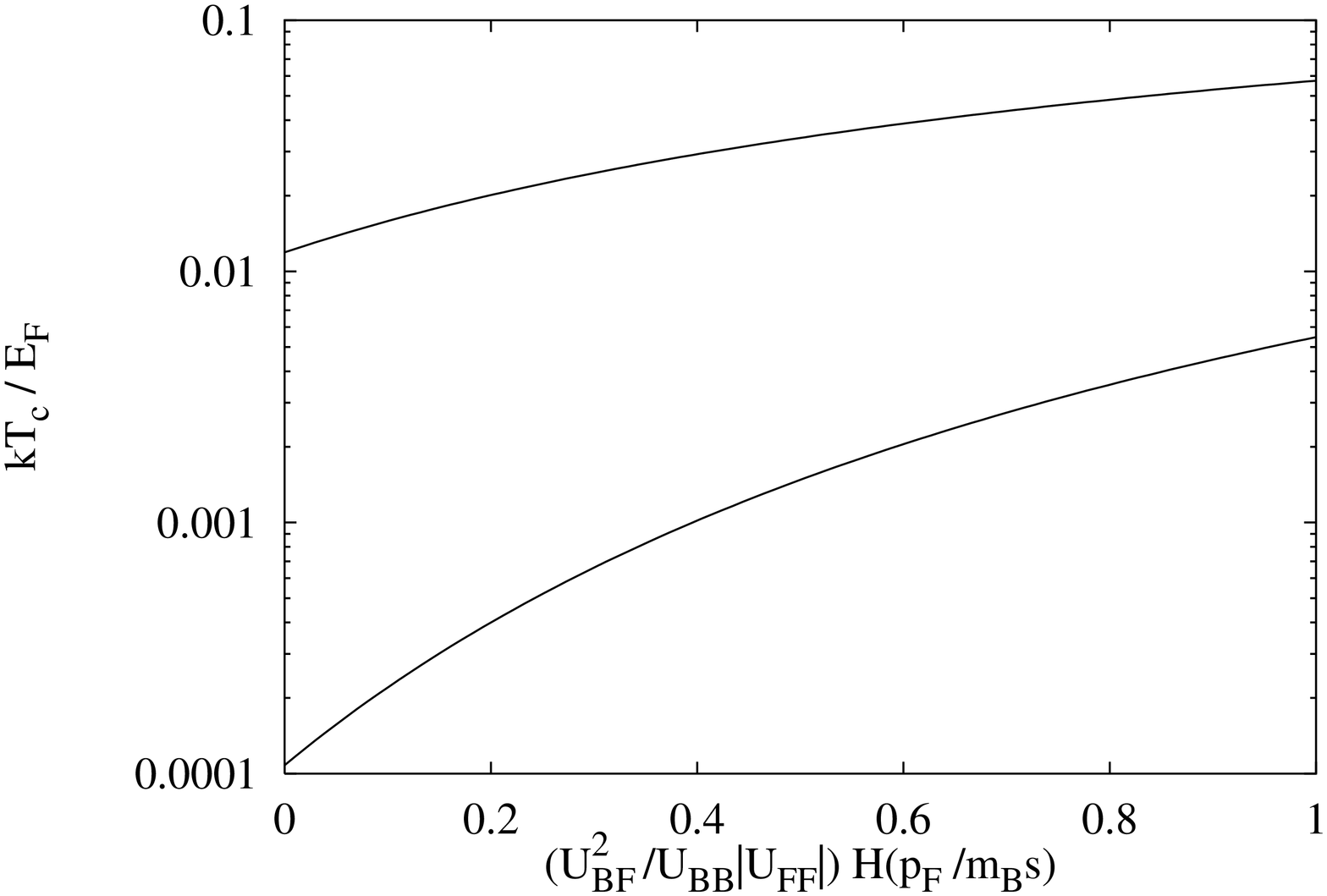,height=7.5cm,angle=-90}
\vspace{.2cm}
\begin{caption}
{Transition temperature as a function of the parameter
$(U_{BF}^2/U_{BB}
|U_{FF}|)H(p_F/m_Bs)$, for $k_F|a|=0.2$ (lower curve)
and $k_F|a| =0.5$ (upper curve).}
\end{caption}
\end{center}
\label{fig3}
\end{figure}

Let us now examine a specific example, a mixture of the fermionic
atoms $^6$Li with
the bosonic atoms $^{87}$Rb.
For the boson-boson  scattering length we use $a_{BB}=109 a_0$,
$a_0$ being the Bohr radius, and for the
fermion-fermion one  $a=-2160 a_0$.
The boson-fermion scattering length is not known.
If we choose the order-of-magnitude estimate
$|a_{BF}|=100 a_0$ 
\cite{stoof},  $U_{BF}^2/U_{BB}|U_{FF}|$ is only 0.18. 
For $n_B= 10^{14}$ cm$^{-3}$ the 
increase in $T_c$ then amounts to factors of 2.7 and 1.3 for  
$k_F|a| =0.2$ and $k_F|a| =0.5$, respectively. If however $|a_{BF}|$ were
twice as large, these factors would change to 17.6 and 2.2. This indicates the
importance of obtaining information about boson-fermion scattering lengths.

Finally we remark that more detailed calculations are required to take into 
account induced interactions when they become comparable in magnitude to the 
bare interaction. The details of the frequency dependence of the induced 
interactions will then become important.  We note that corrections to the 
effective mass of a fermion and to the renormalization constant give 
corrections to $T_c$ of higher order than those considered here.

We are grateful to H.\ T.\ C.\ Stoof for useful conversations.  
LV acknowledges support from the Danish Ministry of Education.

\end{document}